\documentclass[useAMS,usenatbib]{mn2e}
\usepackage[dvips]{graphicx}
\usepackage{longtable}
\usepackage{epsfig}
\usepackage[colorlinks,urlcolor=blue]{hyperref}
\usepackage{amssymb,amsfonts,amsmath,mathtext,cite,enumerate,float}
\usepackage{xcolor}
\usepackage{multirow}
\usepackage{textcomp}

\title[The optical identification of events with poorly defined locations: \textit{Fermi} GBM GRB~140801A]{\large{The optical identification of events with poorly defined locations: The case of the \textit{Fermi} GBM GRB~140801A}}
\author[V. M. Lipunov et al.]{\small{V. M. Lipunov$^{1,2,3}$\thanks{E-mail: lipunov2007@gmail.com (VML)}, J. Gorosabel$^{4,5,6}$\thanks{deceased}, M. V. Pruzhinskaya$^{2}$, A. de Ugarte Postigo$^{4,7}$, V. Pelassa$^{8,29}$, A. E. Tsvetkova$^{9}$,  I. V. Sokolov$^{10}$},
\newauthor \small{D.A. Kann$^{11}$, Dong Xu$^{7,12,28}$, E. S. Gorbovskoy$^{2,3}$, V. V. Krushinski$^{13}$, V. G. Kornilov$^{1,2}$, P. V. Balanutsa$^{2}$, S. V. Boronina$^{22}$, N. M. Budnev$^{14}$, }
 \newauthor \small{Z. Cano$^{15}$, A. J. Castro-Tirado$^{4}$, V. V. Chazov$^{2}$, V. Connaughton$^{8,30}$, C. Delvaux$^{17}$, D. D. Frederiks$^{9}$, J. F. U. Fynbo$^{7}$, A. V. Gabovich$^{18}$,}
   \newauthor \small{A. Goldstein$^{19}$,   J. Greiner$^{17,20}$, O. A. Gress$^{14}$, K. I. Ivanov$^{14}$, P. Jakobsson$^{15}$, S. Klose$^{11}$, F. Knust$^{17}$, V. N. Komarova$^{21}$, E. Konstantinov$^{14}$,}
     \newauthor \small{A. V. Krylov$^{2}$, D. A. Kuvshinov$^{1,2}$,A. S. Kuznetsov$^{2}$, G.V. Lipunova$^{2}$, A. S. Moskvitin$^{21}$, V. D. Pal'shin$^{9}$, S. B. Pandey$^{23}$, V. A. Poleshchuk$^{14}$,}
       \newauthor \small{S. Schmidl$^{11}$, Yu. P. Sergienko$^{18}$, E. V. Sinyakov$^{18}$, S. Schulze$^{24,25}$, V. V. Sokolov$^{21}$, T. N. Sokolova$^{21}$, M. Sparre$^{7}$,  C. C. Th\"{o}ne$^{4}$, A. G. Tlatov$^{26}$,}
       \newauthor \small{ N. V. Tyurina$^{2}$, M. V. Ulanov$^{9,27}$, S. A. Yazev$^{14}$, V. V. Yurkov$^{18}$}\\
$^{1}$M.V.Lomonosov Moscow State University, Faculty of Physics, Leninskie gory, GSP-1, Moscow, 119991, Russia \\
$^{2}$M.V.Lomonosov Moscow State University, Sternberg Astronomical Institute, Universitetsky~pr.,~13,~Moscow, 119234, Russia\\
$^{3}$``Extreme Universe Laboratory'' of Lomonosov Moscow State University Skobeltsyn Institute of Nuclear Physics,\\ 1(2), Leninskie gory, GSP-1, Moscow, 119991, Russia\\
$^{4}$Instituto de Astrof\'{\i}sica de Andaluc\'{\i}a, Glorieta de la Astronom\'{\i}a s/n, Granada,18008 Spain\\
$^{5}$Ikerbasque, Basque Foundation for Science, Alameda de Urquijo 36-5, E-48008 Bilbao, Spain\\
$^{6}$Unidad Asociada Grupo Ciencias Planetarias UPV/EHU-IAA/CSIC, Departamento de F\'{\i}sica Aplicada I, E.T.S. Ingenier\'{\i}a, \\
Universidad del Pa\'{\i}s-Vasco UPV/EHU, Alameda de Urquijo s/n, E-48013 Bilbao, Spain\\
$^{7}$Dark Cosmology Centre, Niels-Bohr-Institute, University of Copenhagen, Juliane Maries Vej 30, DK-2100 K$\o$benhavn $\O$, Denmark\\
$^{8}$Center for Space Plasma and Aeronomic Research (CSPAR), University of Alabama in Huntsville, Huntsville, AL 35899, USA\\
$^{9}$Ioffe Institute, Polytechnicheskaya 26, St. Petersburg, 194021, Russia\\
$^{10}$Terskol Branch of Institite of Astronomy of the Russian AS, Tyrnyauz, 361623, Russia \\
$^{11}$Th\"uringer Landessternwarte Tautenburg, Sternwarte 5,  07778 Tautenburg, Germany\\
$^{12}$National Astronomical Observatories, Chinese Academy of Sciences, Beijing, 100012, China\\
$^{13}$Kourovka Astronomical Observatory, Ural Federal University, Lenin ave. 51, Ekaterinburg 620000, Russia\\
$^{14}$Applied Physics Institute, Irkutsk State University, 20, Gagarin blvd,664003, Irkutsk, Russia\\
$^{15}$Centre for Astrophysics and Cosmology, Science Institute, University of Iceland, Dunhagi 5, IS-107 Reykjavik, Iceland\\
$^{16}$Unidad Asociada Departamento de Ingenier\'{\i}ia de Sistemas y Autom\'{\i}atica, Universidad de M\'{\i}alaga, Spain\\
$^{17}$Max-Planck Institute for Extraterrestrial Physics, Giessenbachstrasse, 85748 Garching, Germany\\
$^{18}$Blagoveschensk State Pedagogical University, Lenin str., 104, Amur Region, Blagoveschensk 675000, Russia\\
$^{19}$Space Science Office, VP62, NASA/Marshall Space Flight Center Huntsville, AL 35812, USA\\
$^{20}$Excellence Cluster Universe, Technische Universit\"{a}t M\"{u}nchen,  Boltzmannstra{\ss}e 2, 85748 Garching, Germany\\
$^{21}$Special Astrophysical Observatory of the Russian AS, Nizhnij Arkhyz 369167, Russia\\
$^{22}$St. Petersburg State University 7-9, Universitetskaya nab., St. Petersburg, 199034, Russia\\
$^{23}$Aryabhatta Research Institute of Observational Sciences, Manora Peak, Nainital - 263 002, India\\
$^{24}$Instituto de Astrof\'{i}sica, Facultad de F\'{i}sica, Pontificia Universidad Cat\'{o}lica de Chile, Vicu\~{n}a Mackenna 4860, 7820436 Macul, Santiago, Chile\\
$^{25}$Millennium Institute of Astrophysics, Vicu\~{n}a Mackenna 4860, 7820436 Macul, Santiago, Chile\\
$^{26}$Kislovodsk Solar Station of the Main (Pulkovo) Observatory RAS, P.O.Box 45, ul. Gagarina 100, Kislovodsk 357700, Russia\\
$^{27}$Peter the Great St. Petersburg Polytechnic University, Polytechnicheskaya 29, St. Petersburg, 195251, Russia\\
$^{28}$Key Laboratory of Space Astronomy and Technology, National Astronomical Observatories, Chinese Academy of Sciences, 20A Datun Road, \\
Beijing 100012, China\\
$^{29}$Fred Lawrence Whipple Observatory, Harvard-Smithsonian Center for Astrophysics, Amado, AZ 85645, USA\\
$^{30}$Universities Space Research Association, Huntsville, AL 35805, USA}
\begin{document}

\date{Accepted 2015 XXX. Received 2015 June; in original form 2015 June}
\pagerange{\pageref{firstpage}--\pageref{lastpage}} \pubyear{2015}
\maketitle
\label{firstpage}

\begin{abstract}
\footnotesize{We report the early discovery of the optical afterglow of gamma-ray burst (GRB)\,140801A in the 137~deg$^2$ 3-$\sigma$ error-box of the \textit{Fermi} Gamma-ray Burst Monitor~(GBM). MASTER is the only  observatory that automatically react to \textit{all} Fermi alerts. GRB~140801A is one of the few GRBs whose optical counterpart was discovered solely from its GBM localization. The optical afterglow of GRB~140801A was found by MASTER Global Robotic Net 53 sec after receiving the alert, making it the fastest optical detection of a GRB from a GBM error-box. Spectroscopy obtained with the 10.4-m Gran Telescopio Canarias and the 6-m BTA of SAO RAS reveals a redshift of $z=1.32$. We performed optical and near-infrared photometry of GRB~140801A using different telescopes with apertures ranging from 0.4-m to 10.4-m. GRB 140801A is a typical burst in many ways. The rest-frame bolometric isotropic energy release and peak energy of the burst is $E_\mathrm{iso} = 5.54_{-0.24}^{+0.26} \times 10^{52}$~erg and $E_\mathrm{p, rest}\simeq280$~keV, respectively, which is consistent with the Amati relation. The absence of a jet break in the optical light curve provides a lower limit on the half-opening angle of the jet $\theta=6.1$ deg. The observed $E_\mathrm{peak}$ is consistent with the limit derived from the Ghirlanda relation. The joint \textit{Fermi} GBM and Konus-\textit{Wind} analysis shows that GRB\,140801A could belong to the class of intermediate duration. The rapid detection of the optical counterpart of GRB\,140801A is especially important regarding the upcoming experiments with large coordinate error-box areas.}
\end{abstract}


\section{Introduction}
A new generation of experiments that will open new observational windows for the study of the Universe is beginning to come online, or will do so in the next years. They will detect gravitational waves (LIGO/VIRGO, etc.), neutrinos (ANTARES, IceCube), observe the Universe's most energetic particles (MAGIC, HESS, CTA, HAWC, etc.) or the low frequency end of the radio spectrum (LOFAR, MeerKAT, SKA, etc.). Their common characteristic is that they will detect short-lived transient objects, which will probably have optical counterparts, but with localization error-boxes that are too large to allow direct follow-up from most large optical telescopes. Being able to rapidly detect the optical transients with wide-field robotic observatories is the key to allow observations with large optical telescopes that will characterize their progenitors.

In fact, the current situation in gravitational wave, radio and neutrino astronomy resembles one that took place during 30 years (until 1997) in gamma-ray burst (GRB) astronomy. However, now things have  changed  dramatically due to the emergence of wide-field robotic telescopes with excellent software, allowing us to discover new objects in wide-field sky regions almost in real time. In addition, the number of big telescopes has grown significantly over the past few decades. A special place among these telescopes belongs to the biggest optical telescope in the world --- the 10.4-m Gran Telescopio Canarias (GTC).

Observational astronomy is significantly challenged by the Gamma-Ray Burst
Monitor (GBM) on the Fermi gamma-ray space observatory, whose position
error-boxes reach hundreds of square degrees. Undoubtedly, the study of GRBs remains a very important astrophysical problem in itself. Thus, collective observations of gamma-ray bursts, representing a new global astrophysical experiment, provide an excellent opportunity for developing new software dedicated to optical-transient (OT) detection in  wide-field sky regions.

In this paper we demonstrate the cooperation of space, ground-based, robotic and semi-automatic systems in the context of the optical identification of GRB~140801A discovered by the \textit{Fermi} Gamma-ray Burst Monitor (GBM) on 2014 August 1 (FERMI trig. 428612396;~\citealt{16653}). MASTER is the unique system, that react to \textbf{all} \textit{Fermi} alerts (fully automatically). The afterglow of GRB~140801A, discovered by MASTER \citep{Lipunov2010}, is one of the few discovered just with a \textit{Fermi} GBM alert. The first afterglow discovered solely based on a \textit{Fermi} GBM localization was made by the intermediate Palomar Transient Factory (iPTF) for GRB~130702A~\citep{iPTF}. With a first detection centred just 110~s
after the GRB trigger, the OT of GRB~140801A was detected by MASTER much more rapidly than any detections from iPTF~\citep{Singer2015}.

The paper is organized as follows. The MASTER observational strategy for large error-boxes is given in Section 2.  We describe the discovery timeline of GRB~140801A, present our observations, photometry, and spectroscopy in Section 3.  A discussion of the characteristics of the GRB,
its afterglow and its host galaxy is given in Section~4. We conclude the paper in Section~5.

\section{MASTER observing strategy}
The MASTER Global Robotic Net\footnote{\url{http://observ.pereplet.ru/}} includes six observatories with identical instruments: MASTER-Amur,  MASTER-Tunka, MASTER-Ural, MASTER-Kislovodsk (Russian Federation), MASTER-SAAO (South Africa) and MASTER-IAC (Spain, Canarias), and Very Wide Field cameras (MASTER-VWF) in Argentina~\citep{Lipunov2004,Lipunov2010,Kornilov2012,Gorbovskoy2013}. Each MASTER observatory  provides a survey speed of 128~deg$^2$ per hour with a limiting magnitude of 20~mag on dark, moonless nights. Each observatory is equipped with a twin-tube aperture system with a total field of view of 8~deg$^2$,  with a 4098~pixel~$\times$~4098~pixel CCD camera with a scale of 1.85$''$/pix, an identical MASTER photometer with \textit{BVRI} filters. It is possible to observe without any filter in integral (white) light with or without two orthogonal polarizers. Details of the MASTER filters and polarization measurements can be found in \citet{Kornilov2012,Gorbovskoy2013}.

The observations with MASTER-Net can be performed in different modes: alert, survey, and inspection. The alert mode is on if a target position has good accuracy (when the error-box size is less than  2$^\circ\times$2$^\circ$ MASTER FOV)  and is usually used to observe GRBs upon receiving notices from the Gamma-ray Coordinates Network\footnote{\url{http://gcn.gsfc.nasa.gov/}} (GCN).
 In the alert mode MASTER observes with parallel tubes and with polarizers. Exposure times follow the relation $t_{exp} = (T_{start}-T_{0})/5$, where $T_{0}$ is the trigger time (UT), $T_{start}$ is the time of the beginning of exposure (UT). The exposure time is rounded to an integer with a step of 10 s and cannot exceed 3 min.

The survey mode is used for the regular survey and search for optical transients (OTs) when there are no GRB alerts.
MASTER software and the planner has been developed to select  preferred locations for the survey. The planner takes into account the previous coverage rate of the area, distances from the Galactic plane, the Moon, the Sun, the ecliptic, and the current \textit{Swift} FOV. It takes into account the number of SNe~Ia in the field and GRBs  discovered in the vicinity in the previous 24  hours.
During the survey each area is observed several times with automatically chosen exposure times ranging from 60 to 180~s. The lapse time varies from 10~min to 1~h depending on the Moon phase, weather conditions, and the remaining time of observations.

In case of large coordinate error-box areas (mainly for GRBs from the \textit{Fermi} observatory), we use the inspection mode, which combines the alert and survey modes. Firstly, the centre of the error-box is observed in the alert mode during the time  $t-T_{0} < 5$ min. Then the telescope switches to the survey mode inside the error-box area. The 1-$\sigma$ error-box is covered first, then 2-$\sigma$ and 3-$\sigma$. The error-boxes are covered using the same algorithm as for the survey mode. Each area is observed three times with five minute intervals with exposure times 60~s. The inspection mode allows us to cover big areas quickly and search for all types of OTs.  If the same error-box can be observed by two or more telescopes of MASTER-net, they will cover different fields. Thus, the rate of coverage will grow in proportion to the number of telescopes.

In case of a modification of the error-box position, alert observations are performed for each adjustment. For inspection mode the smallest error-box is chosen.

The main MASTER unique feature is own software,  which was written over the course of 10 years, and which
has let us to discover new optical transients in MASTER images within several minutes after readout from the CCD. This information includes the full classification of all sources from the image, the data from previous MASTER-Net archive images for every sources, full information from VIZIER database and all open source (e.g., Minor planet checker center), derivation of orbital elements for moving objects and more.
 In search tasks the real astrophysical sources are not likely to be represented by just 1, 2 or 4 pixels in images, such sources are very likely to be
artificial and are screened out by the search task.  Real transient sources must have more then 10 pixels distributed with a specfic
profile to distinguish them from hot pixels. The MASTER software discovers optical transients not by the difference between the
previous and current images, but by the fully identifying each source in every image.

The MASTER software gives us full information about all optical sources in every image, including astrometry, photometry, the parameters for moving objects and the possibility to compare with archive MASTER images and with all public data. This MASTER software has allowed us to discover  900 optical transients of ten different types (GRB optical counterparts, supernovae (including the superluminous one), novae, QSO flares, short transients, dwarf novae, antinovae($\varepsilon - Aur type$), RCB and other cataclysmic variables, UVCet-type stars, potentially hazardous asteroids and comets) \footnote{\url{http://observ.pereplet.ru/MASTER_OT.html}}.

The discovery strategy for optical transients is the
following.

The image processing (the key steps are described in Section 8, \citet{Kornilov2012}) is started after the image has been trasmitted from the CCD to
the database. The alert images have the highest priority for processing. The first stage is the primary data reduction including the bias and dark frame subtraction and flat-fielding. The next stage is object detection and determination of their coordinates. As a base for object detection and classification, the SExtractor package \footnote{\url{http://sextractor.sourceforge.net/}}  developed for the TERAPIX project was adopted and additionally tuned for our images application. The table of detected objects is used for the sky patch identification and calculation of the coordinates conversion coefficients. The photometric reduction is made using minimization of differences between the instrumental and catalogue magnitudes for all the reference stars selected on the current frame.

The detected objects (at MASTER image) are classified into three categories:
1. the known stars — the objects are identified by mathcing its coordinates and magnitude to the catalogue;
2. $"$flare$"$ —  the object is at the same coordinates as an object from the catalogue but has a significant difference in magnitude (either negative or positive);
3. unknown – the object is absent from the catalogues.
 Then we compare the object lists to filter out uncatalogued moving objects and to start work with found transient.

This is a fully automatically detection system and takes 1-2 minutes from CCD readout. After software's automated OT detection and primary classification, each candidate is carefully analysed by humans to further investigate its nature.

If we have several images with the OT, we analyse its light curve (LC) and MASTER archive images. Then a human also analyses the public databases (like VIZIER) in this area. If the error box of an alert has been imaged, and there are no sources in previous or archived MASTER images (lists of objects),
and no sources inside 5" in the VIZIER database, this is likely the
optical counterpart of a GRB.
If there is a galaxy in the neighbourhood (the software automatically checks this and classifies the OT as a PSN), we can classify the OT as a possible SN (after manual checking of this position to find any faint (less then the image optical limit) Milky Way's source in the sight beam in MASTER or POSS archive images). If we analyze MASTER's Andromeda image~\footnote{\url{http://observ.pereplet.ru/MASTER-M31.jpg}}, we can tell about possible Nova in M31.  If there is no any VIZIER sources inside 5" and
the LC is constant over 1-2 nights, it can be a cataclysmic variable
(mostly of the dwarf nova type). If the LC increases and fades away
again over the course of several tens of minutes and there is a red
or infrared detection in VIZIER, it is likely to be a UV Cet type
object.

\section{GRB~140801A observations}
\subsection{MASTER optical counterpart discovery}

MASTER is the single observatory that automatically react to all Fermi alerts.

On 2014 August 1 at 18:59:53.26 UT, the \textit{Fermi} GBM detected trigger 428612396. The MASTER-Net was the first ground-based facility to find the optical counterpart of GRB~140801A, within a big field of view, i.e. 137 square degree error-box.

After receiving  the alert, MASTER-Tunka began an automatic pointing towards the centre of the \textit{Fermi} GBM error-box 54s after notice time~\citep{16653} (alert-mode). During the slew, MASTER received  notices that modified the error-box position twice and had to repoint. Thus, MASTER started to observe the error-box centred on $\alpha=03^h$ 01$^m$ 00$^s$, $\delta=+30^\circ$ 53$'$ 24$''$ (J2000) with 1-$\sigma$ error radius of 1.2$^\circ$ (statistical uncertainty only). MASTER obtained 5 images of this region with increasing  exposure time in white light and polarizers. And then MASTER-Tunka starts to cover Fermi error-box in inspect mode and covered 52 sq.degrees.

The MASTER real-time auto-detection system found 81 new and variable optical transients in this region. 18 known asteroids, 12 mooving objects, 4 hot pixels and  14 bright star's patches of light were discriminated automatically. The remaining 22 faint stars, 10 parts of faint galaxies and 1 real transient were inspected manually.

\begin{figure*}
\centering
\center{\includegraphics[width=1\linewidth]{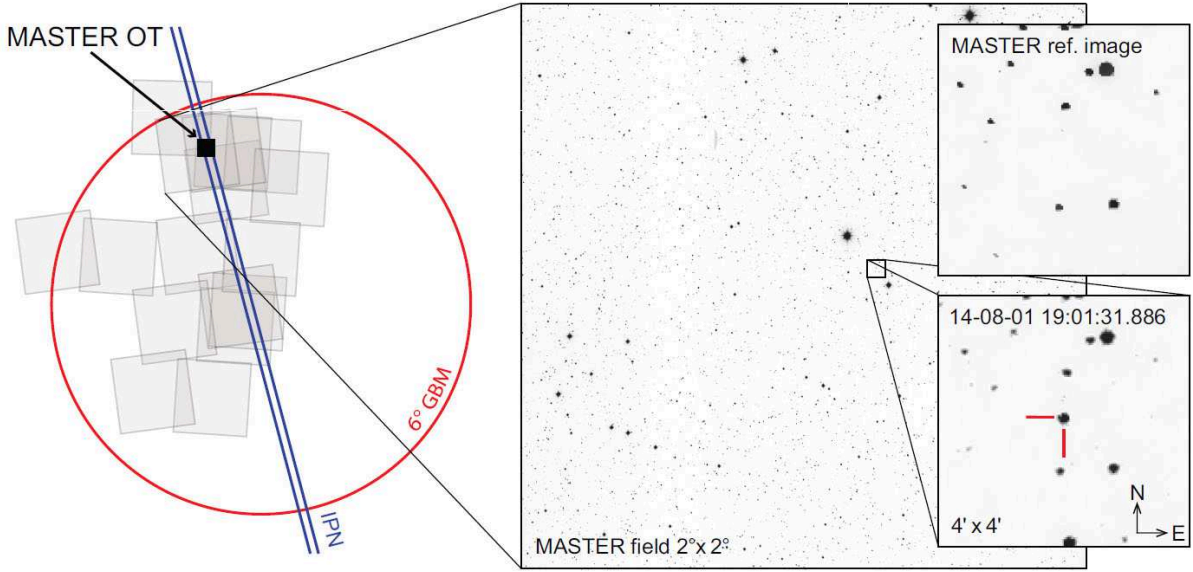}}
\caption{MASTER OT discovery of GRB~140801A in the \textit{Fermi} GBM error-box. The left panel illustrates the final GRB localizations (red circle: 3$^{\circ}$ of statistical error (3-$\sigma)$ and $3^{\circ}$ of systematic error GBM; blue lines: 3-$\sigma$  IPN). Gray squares are fields covered by MASTER. The black square is the location of the MASTER OT. The right panels show MASTER images of the OT position: discovery image (bottom), reference image (top). The IPN localization was published after the MASTER discovery circular~\citep{GCN16655,16653}.}
\label{Position}
\end{figure*}

A bright OT source  at $\alpha=02^h$ 56$^m$ 16.44$^s$,  $\delta=+30^\circ$ 56$'$ 16.8$''$ (J2000) was identified as the possible optical afterglow of GRB~140801A~\citep{16653}. The first image with the GRB candidate began 54 s after receiving the alert and 99 s after the trigger time. The OT magnitude in white light was 14.64$\pm$0.07~mag. The OT was also clearly seen in the next 4 images in polarized light. On August 1 at 19:57:26.72 UT MASTER returned to the OT position in inspection mode and found nothing to a 3-$\sigma$ upper limit of 19.82~mag. We would like to stress that the IPN triangulation, \textit{Fermi} and \textit{Swift} XRT circulars, reporting refined localizations, appeared after the MASTER OT detection (see Fig.~\ref{Position}). And \textit{Swift} observations of the X-ray afterglow were the result of MASTER OT discovery.
At the other sites of MASTER-Net observations were not
performed due to several reasons: bad weather conditions (MASTER-Ural), sunrise (MASTER-Amur) and object being below the horizon at the moment of the alert (MASTER-Kislovodsk).

\begin{figure*}
\centering
\center{\includegraphics[width=1\linewidth]{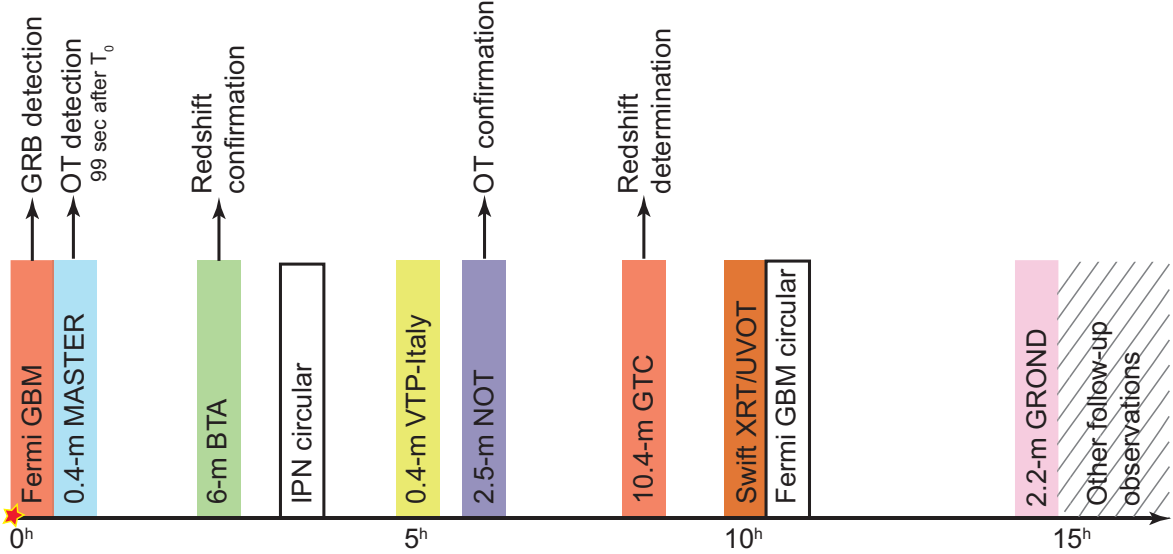}}
\caption{Observational timeline. The figure shows the approximate time of observations obtained for GRB~140801A by the different telescopes. For IPN and \textit{Fermi} GBM the time of GCN circular publication  is shown. \textit{Fermi} GBM --- \textit{Fermi} Gamma-ray Burst Monitor~\citep{GCN16658}; 0.4-m MASTER ---  Global  MASTER  Robotic  Net telescope located in Tunka~\citep{16653}; 6-m BTA ---  Big Telescope Alt-azimuth of the Special Astrophysical Observatory of the Russian Academy of Sciences~\citep{GCN16663}; IPN ---  InterPlanetary Network~\citep{GCN16655}; 0.4-m VTP-Italy --- telescope of the Virtual Telescope Project~\citep{GCN16659}; 2.5-m NOT --- Nordic Optical Telescope~\citep{GCN16656}; 10.4-m GTC ---   Gran Telescopio Canarias~\citep{GCN16657}; \textit{Swift} XRT/UVOT --- \textit{Swift} X-ray Telescope and Ultraviolet/Optical Telescope~\citep{GCN16661,GCN16662}; 2.2-m GROND --- Gamma-Ray Burst Optical/Near-Infrared Detector mounted at the 2.2-m MPG telescope at ESO La Silla Observatory~\citep{
GCN16666}.}
\label{Schedule}
\end{figure*}

MASTER published a GCN circular on August 1 at 20:52:09 GMT and, therefore, became the first  telescope to report the  optical counterpart of GRB~140801A~\citep{16653}. The subsequent Inter Planetary Network (IPN) triangulation showed that the MASTER OT position lies within the 3-$\sigma$ position annulus, and is therefore most probably the optical afterglow of the gamma-ray burst~\citep{GCN16655}. The next telescope that confirmed the OT  at the MASTER position was the 2.5-m Nordic Optical Telescope (NOT)~\citep{GCN16656}. The OT redshift was determined by the 10.4-m GTC and later confirmed by the 6-m Big Telescope Alt-azimuth of SAO RAS (BTA)~\citep{GCN16657,GCN16663}. The OT was also observed by different optical telescopes and further analysis was reported~\citep{GCN16659,GCN16660,GCN16661,GCN16666},etc. The observational timeline is presented in Fig.~\ref{Schedule}.

\subsection{High-energy observations}
\begin{figure*}
\center
\includegraphics[width=0.8\textwidth]{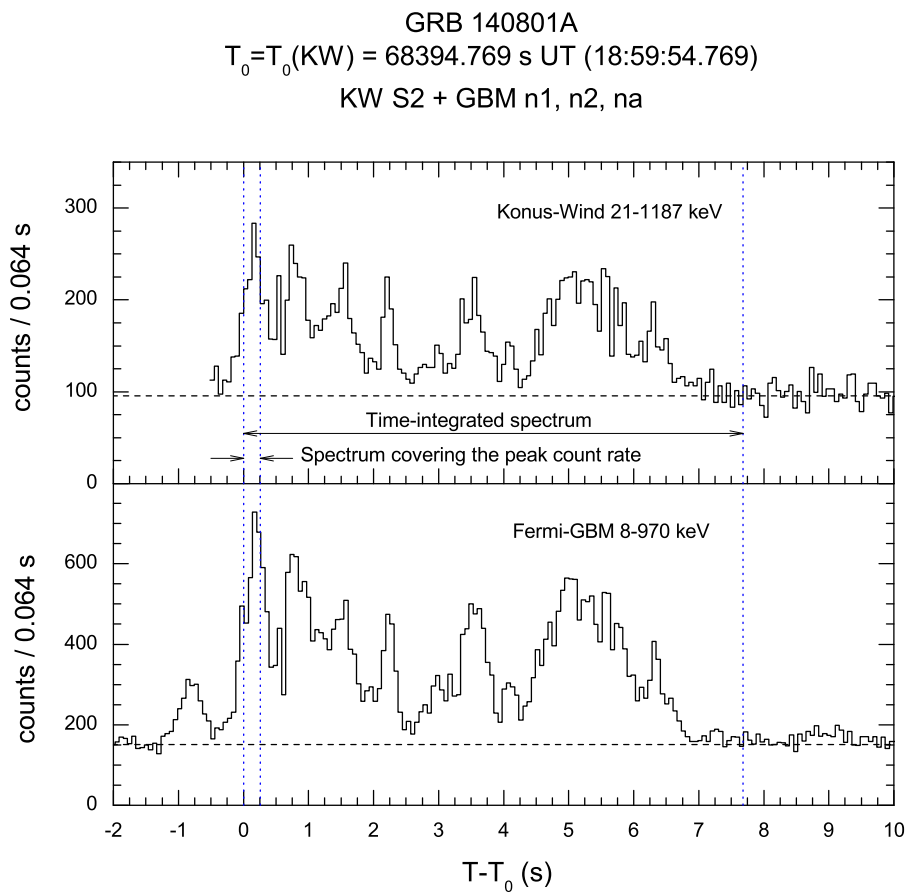}
\caption{
KW (21--1187~keV, top panel) and \textit{Fermi}~GBM (8--970~keV, bottom panel) light curves with 64~ms resolution.
The data are plotted relative to the KW $T_0=18$:59:54.769~UT. The GBM light curve is shifted by 0.639~s to account for the propagation delay between \emph{Fermi} and \emph{Wind}.
The accumulation intervals for the time-integrated spectrum and the spectrum covering the peak count rate are indicated by vertical dashed lines; the dashed horizontal lines denote the background level.
}
\label{figKWGBMlc}
\end{figure*}

The {\it Fermi} GBM (\citealt{Meegan09}) detected and localized GRB 140801A at 18:59:53.26 UT \citep{GCN16658}. The final localization from GBM data, published as a GCN notice 1 h 6 min after the trigger, is $\alpha = 45.8^{\circ}$, $\delta = +26.4^{\circ}$ with a $1^{\circ}$ error radius (68~per cent, statistical uncertainty only). The typical systematic error of a GBM localization is 2 to 3$^{\circ}$ \citep{Connaughton15}. The first GBM on-flight localization, distributed as a GCN notice, had an error radius of $3.85^{\circ}$ (68~per cent, statistical uncertainty only). The GBM NaI light curve in the 8--970 keV band is shown in Fig.~\ref{figKWGBMlc} (bottom panel). We wrote a python script to reduce the TTE data and apply the propagation delay from \textit{Wind} to \textit{Fermi} to them. The burst duration measured by GBM is $T_{90} = 7.2 \pm 0.6$ s in the 50--300 keV energy range (GBM catalog; \citealt{vonKienlin14}). The data used in the spectral analysis are from the NaI detectors 1, 2 and 10 (8--1000~keV) which saw the GRB at an angle of less than 60$^{\circ}$ and are not blocked by the satellite, solar panels, or the {\it Fermi} Large Area Telescope, and both BGO detectors (150~keV -- 40~MeV). \textsc{xspec12}~\citep{Arnaud1996} compliant PHA and BAK files are extracted from the GBM TTE datafiles using the \textsc{rmfit} analysis software\footnote{\url{http://fermi.gsfc.nasa.gov/ssc/data/analysis/rmfit/}}.

\begin{figure*}
\center{\includegraphics[trim=0 230 0 0,clip, width=0.8\textwidth]{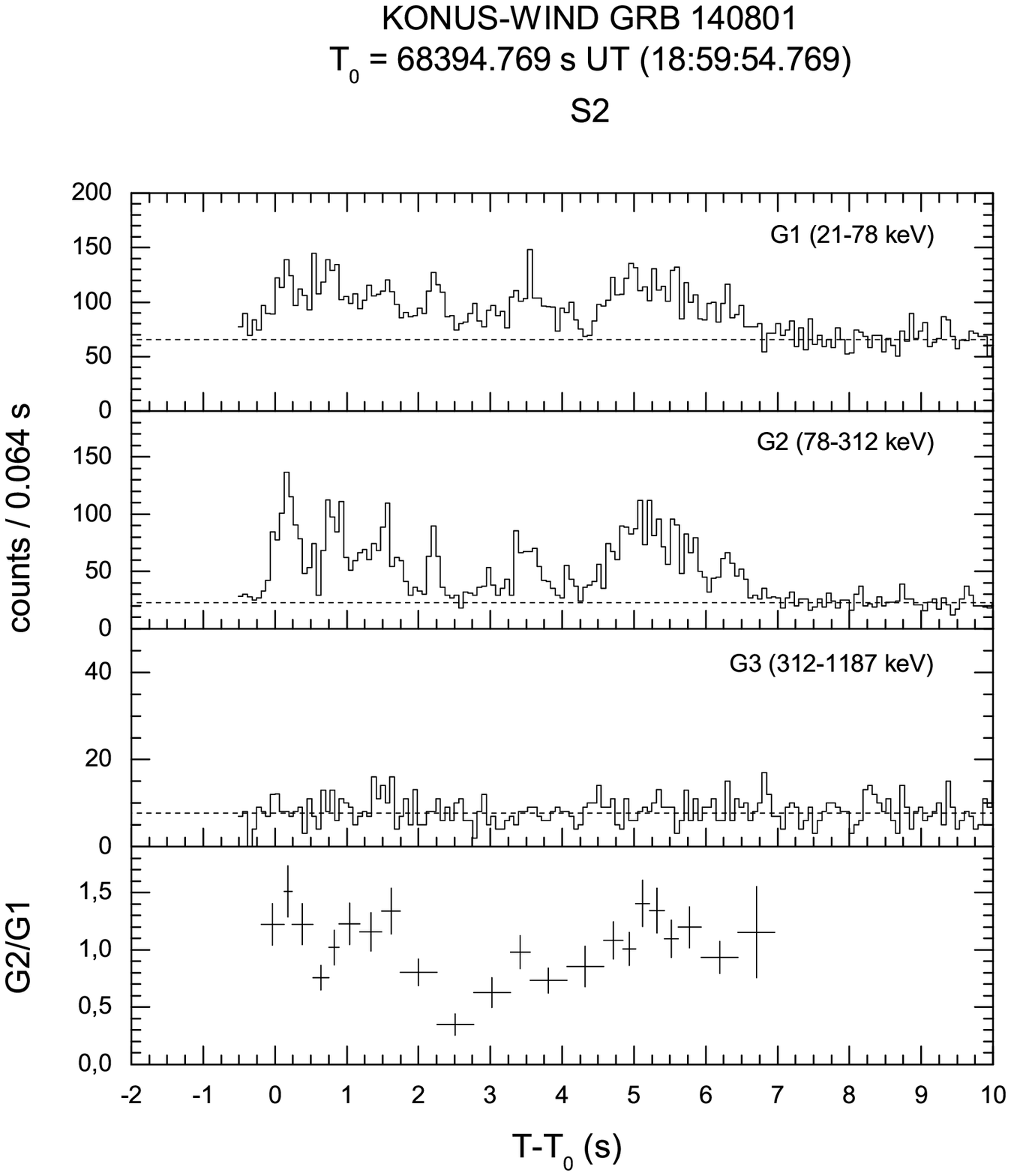}}
\caption{
Light curve of GRB~140801A recorded by Konus-\textit{Wind} in the G1, G2, and G3 energy bands with 64~ms resolution (three upper panels) and the G2/G1 hardness ratio (lower panel). The count rates are dead-time corrected, and background levels are indicated by dashed lines.
}
\label{figKWlc}
\end{figure*}

GRB~140801A triggered the S2 detector of the Konus-{\it Wind} (KW) gamma-ray burst spectrometer (\citealt{Aptekar1995}) at $T_0$=18:59:54.769~UT~\citep{GCN16660}. Fig.~\ref{figKWlc} shows the KW light curve in three energy bands: G1 (21--78~keV), G2 (78--312~keV), and G3 (312--1187~keV).
The light curve has a multi-peaked structure with a total duration, determined at the 5-$\sigma$ level, $T_{100} = 8.03$~s in the G2 band;
the corresponding $T_{90}$ value is $6.2_{-0.5}^{+0.2}$~s, which is consistent with the GBM duration (all the errors quoted in this section are at the 90 per cent confidence level).
As it can be seen from Fig.~\ref{figKWlc}, the burst emission above $\sim$\,300~keV (the G3 band) is virtually indistinguishable from the background. Shown in the same Figure, the temporal evolution of the G2/G1 hardness ratio is consistent with the hardness-intensity correlation~\citep{Golenetskii1983}.
The spectral lag between the light curves in the G2 and G1 bands is examined using the cross-correlation function ~\citep{Band1993}.
The resulting lag value $\tau_{lag} = -0.01 \pm 0.03$~s suggests that the observed spectral lag is negligible.

A spectral analysis has been performed using the KW and GBM (NaI and BGO) data for two time intervals:
from $T_0$ to $T_0+7.680$~s, the time-integrated spectrum;
and from $T_0$ to $T_0+0.256$~s, the spectrum covering the peak count rate. The KW spectral data cover the 20~keV -- 15~MeV energy range (the emission is seen up to $\sim$\,3~MeV).
Fig.~\ref{figKWGBMlc} shows the KW and GBM light curves with the spectra accumulation intervals indicated.
Results of the KW, GBM, and joint KW-GBM spectral fits obtained with \textsc{xspec12}~\citep{Arnaud1996} are summarized in Table~\ref{fittab}.
Two spectral models were tested, the exponential cutoff power-law (CPL), and the Band GRB function (\citealt{Band1993}), both normalized to the energy flux in the observer-frame 1 keV -- 10 MeV range.
In the joint fits, the models were multiplied by a normalization factor to take into account the systematic effective area uncertainties in the response matrices of each instrument.
The normalization factor of the KW data was fixed to 1, and the normalization factors of the GBM spectra were the free parameters.
We found that the results of the KW, GBM, and joint KW-GBM spectral fits are in reasonable agreement; in the latter case, the GBM-NaI normalization factor was found to be close to unity ($\sim$\,0.96).
As it can be seen from Table~\ref{fittab}, both spectra are well fitted by the CPL model with $E_\mathrm{peak}$ of $\sim$\,110--130~keV and a rather hard low-energy photon index $\alpha$. The low-energy photon indices of the time-integrated and the "peak flux" spectra ($\sim$−-0.23 and $\sim$+0.11, respectively) belong to the hardest 3–-5$\%$ of the $\alpha$ values presented in the \textit{Fermi} GBM Burst Catalog 4 (Gruber et al. 2014; von Kienlin et al. 2014) and in the KW catalog of GRBs with known redshifts (Tsvetkova et al. in prep.). The Band function does not improve the fits, yielding the same values of $E_\mathrm{peak}$ and $\alpha$ with only an upper limit on the high-energy index $\beta$.

From the KW temporal and spectral analysis, the burst fluence from $T_0-1.422$ s  to $T_0+6.608$ s (the $T_{100}$ interval) is 1.17$_{-0.05}^{+0.05} \times 10^{-5}$ erg cm$^{-2}$, and the 64-ms peak flux, measured from $T_0+0.146$ s, is 4.76$_{-0.95}^{+0.97} \times 10^{-6}$ erg~cm$^{-2}$~s$^{-1}$ (both calculated in the observer-frame 1 keV -- 10 MeV energy range).
For the redshift $z=1.32$ \citep{GCN16657,GCN16663} and a standard cosmology model with $H_0 = 67.3$~km~s$^{-1}$~Mpc$^{-1}$ and $\Omega_M = 0.315$~\citep{Plank2014} the isotropic rest-frame energy release and the isotropic rest-frame peak luminosity of the burst in the bolometric energy range ($\sim 2.3~$keV -- $\sim 23~MeV$, rest-frame) are $E_\mathrm{iso} = 5.54_{-0.24}^{+0.26} \times 10^{52}$~erg and $L_\mathrm{peak,iso} = 5.24_{-1.05}^{+1.07} \times 10^{52} \; \rm erg\,s^{-1}$, respectively.

The hardness ratio obtained from the spectral models for the fluences in the 100-300~keV and 50-100 keV bands, $F(100-300)/F(50-100)$, is about 3 and $F(50-100)/F(25-50) \simeq 2.5$. Together with the derived $T_{90}\simeq 6.2$~s this indicates, following the classification of~\citet{Horvath}, that GRB\,140801A could belong to the class of intermediate bursts, although it seems to have border-line spectral hardness  (see Fig.~1 in~\citealt{Horvath}).
GRB\,140801A has a typical redshift for the intermediate group for which the median is $z_\mathrm{med}=1.55$~\citep{Postigo2011}.

\subsection{Optical observations}
From August 1 to November 25 numerous optical and near IR observations were performed by different telescopes. The resulting data are presented in Table~\ref{TabPhot}.
 Observations by the MASTER-Tunka robotic telescope that gave five images in white light
and with polarizers were described above, in Section 3.1. Below, the observations by other facilities are described.

\subsubsection{BTA}
Observation of GRB~140801A with the 6-meter BTA  telescope of SAO RAS\footnote{\url{https://www.sao.ru/Doc-en/index.html}} (Caucasus Mountains, Russia) started two hours after the detection of the burst by the \textit{Fermi} Gamma-ray Space Telescope~\citep{GCN16658}. The GRB field was also observed on the following night (see Table~\ref{TabPhot}).

The observations were carried out with the Scorpio-1 optical reducer\footnote{\url{http://www.sao.ru/hq/lon/SCORPIO/scorpio.html}}~\citep{ScorpioI}, set in the BTA primary focus. Johnson-Cousins broadband $V$  ($\lambda_c=547$~nm, $EW=790$~nm) and $R_c$ ($\lambda_c=662$~nm, $EW=150$~nm)
filters were used for photometry. During the first night a native EEV 42-40 CCD was used, but on the second night due to technical reasons it was replaced with another 42-40 CCD from the Zeiss-1000 telescope. The pixel size of both CCDs is 13.5 \textmu m providing a scale of $0.357^{\prime\prime}$/pix with $2\times2$ binning.

On the first night six $V$ images were obtained, with a total exposure of 180 s, and nine $R_c$ images, with a total exposure of 190 s.
The OT was distinctly seen in individual images.
During the second observational night five $R_c$ images, 180 s each, were obtained.
\subsubsection{NOT}
The 2.5-m Nordic Optical Telescope\footnote{\url{http://www.not.iac.es/}} (Canary Islands, Spain) equipped with ALFOSC started to observe the field of GRB~140801A 6.12 hours after the trigger time. NOT obtained 200~s exposures in  $V$ ($\lambda_c=530$~nm, $FWHM=80$~nm), $R$ ($\lambda_c=650$~nm, $FWHM=130$~nm), and $I$ ($\lambda_c=797$~nm, $FWHM=157$~nm) bands. The OT was clearly detected in each of the images. Observations were again performed 9 days later, but the OT was not found. A faint object of $\sim$\,24~mag was detected on  August 26  in $R$ filter and its magnitude was confirmed by further observations on  November  25 within 1-$\sigma$ error. This indicates that the detection of August 26 is most probably the host galaxy of GRB~140801A.
\subsubsection{GTC}
The 10.4-m Gran Telescopio CANARIAS\footnote{\url{ http://www.gtc.iac.es/}} (Canary Islands, Spain) obtained two images with the OSIRIS camera~\citep{2000SPIE.4008..623C} 8.54 and 9.92 hours after the burst.
Each image consisted of a 20 s exposure obtained through the Sloan broadband $r'$ filter, obtained as acquisition frames for spectroscopy (as reported below).
\subsubsection{GROND}
The Gamma-Ray Burst Optical/Near-Infrared Detector (GROND;~\citealt{Greiner2008}) mounted at the 2.2-m MPG telescope at ESO La Silla Observatory (Chile) performed three  epochs of observations in optical  $g'$ ($\lambda_c= 459$~nm), $r'$ ($\lambda_c=622$~nm), $i'$ ($\lambda_c=764$~nm), $z'$ ($\lambda_c=899$~nm) and NIR $J$ ($\lambda_c=1256$~nm), $H$ ($\lambda_c=1647$~nm) and $K$ ($\lambda_c=2151$~nm) bands~\citep{GCN16666}. The last epoch yielded only upper limits.
\subsubsection{CAHA}
At the Calar Alto Observatory\footnote{\url{http://www.caha.es/}} (Almeria, Spain), 3.5-m telescope equipped with the Omega-2000 near infrared wide field camera obtained two images of GRB~140801A in $H$ ($\lambda_c=1648$~nm, $FWHM=267$~nm) and $K$  ($\lambda_c=2151$~nm, $FWHM =304$~nm)  filters on August 11.

\subsection{Photometry}
The photometric calibration used in this work is based on reference stars in the GROND AB system, which is essentially the SDSS system.  Lupton (2005) equations\footnote{\url{https://www.sdss3.org/dr8/algorithms/sdssUBVRITransform.php}} were applied to calculate \textit{BVRI} magnitudes.
CAHA $H, K$ images were calibrated using 2MASS stars and were corrected for AB offsets (1.39~mag for $H$ and 1.85~mag for $K$ band,~\citealt{Blanton2007}). MASTER $C$ filter (white light) corresponds to $0.2B~+~0.8R$. The details of photometry calibration can be found in~\citet{Gorbovskoy2012}.
All photometry was performed using \textsc{iraf/apphot\footnote{IRAF is distributed by the National Optical Astronomy Observatory, which is operated by the Association of Universities for Research in Astronomy, Inc. under cooperative agreement with the National Science Foundation.}}~\citep{Tody1993}.

The detailed photometric results obtained by the different telescopes are presented in Table~\ref{TabPhot}. The final magnitude errors include the systematic error of the reference stars.
We have not corrected the magnitudes for Galactic extinction. The Galactic extinction $A_V$ is 0.62~mag~\citep{Schlafly2011}.

The optical and XRT 0.3--10 keV light curves are shown in Fig.~\ref{LC}. The XRT data were downloaded from the \textit{Swift} light curve repository~\citep{Evans2007,Evans2009} and were converted to AB magnitudes using formula: $m_\mathrm{AB} = -2.5log_{10}(f_{\nu} [cgs])-48.6$. The OT decays as $t^{-1.42\pm0.04}$ during the MASTER observations, and then, after 2 hours, changes to $t^{-0.81\pm0.04}$ in $V$ and $t^{-0.82\pm0.01}$ in $R$ bands assuming a simple power-law decline ($m = 2.5ax+b$, where $x = log(t)$). We do not see any evidence of a jet break in the optical data. There is a resemblance of GRB 140801A to some previous GRBs in regard of their optical light curve pattern~\citep{Japelj}. Presumably the first steep component can be explained as the reverse shock while the later slowly fading component can be interpreted as the forward shock~\citep{SariPiran,Melandri2008}.

At a distance corresponding to $z = 1.32$, the absolute $R$ magnitude of the host galaxy, obtained from the last NOT $R$ observation,
is estimated to be $\sim$\,$-$20.2~mag. We use a standard cosmology model (see the end of Subsection~3.2), Galactic extinction from~\citet{Schlafly2011} and the relation from~\citet{Wisotzki} for $K$-correction assuming a
spectral index for the host galaxy of $\alpha = -0.5$.

Each tube of each MASTER telescope  is equipped with a polarizer. The polarization directions of polarizers at one site are perpendicular to one another. Thus, one MASTER telescope, when observing alone, is able to measure one Stokes parameter. Details on the Stokes parameter derivation and the calibration of polarization measurements are given in~\citet{Pruzhinskaya2014}.
The differences between the signals obtained by MASTER-Tunka in two polarizers for the time interval from 202.8 to 315.7 s yield the dimensionless time-averaged Stokes parameter $Q = 2.4\pm2.5$~per cent. The error of the Stokes parameter includes the error obtained from the magnitude error and the standard dispersion of the Stokes parameters of the field stars.  The weighted average of the Stokes parameter is calculated for the time interval.  For the derived accuracy of 2.5~per cent, the 1-$\sigma$ upper limit for the degree of linear polarization P is about 18~per cent (see Fig. 14 of~\citet{Gorbovskoy2012}: a value of P = 18~per cent matches 1-$\sigma$ probability $L = 100-68 = 32$~per cent for the curve corresponding to a relative accuracy 2.5~per cent).

\subsection{Spectroscopy}
We obtained spectroscopy of the afterglow of GRB\,140801A  at the 10.4-m Gran Telescopio Canarias~\citep{GCN16657}  and the 6-m BTA telescope of SAO RAS~\citep{GCN16663}.

The GTC observations were performed with the OSIRIS instrument~\citep{2000SPIE.4008..623C} using two grisms: R1000B covering the range between 3630 and 7500 {\AA}
with a resolution of $\sim$\,1000, and R1000R covering the range between
5100 and 10000 {\AA} with a resolution of $\sim$\,1100. Three 900~s exposures were obtained for each of the grisms, with mean times of 8.987 and 10.356~h after the burst, respectively\footnote{The spectral data is  available through GRBspec \url{http://grbspec.iaa.es/}~\citep{2014SPIE.9152E..0BD}}. The data are reduced in
a standard way using appropriate routines based on \textsc{iraf}~\citep{Tody1993}.

The observations at BTA were carried out with the Scorpio-I optical reducer\footnote{\url{http://www.sao.ru/hq/lon/SCORPIO/scorpio.html}}~\citep{ScorpioI},
set in the BTA primary focus. The long-slit spectroscopy was made with the grism VPHG550G covering
the range  between 3500 and 7200 {\AA}. The particular configuration of the device
in combination with the $1^{\prime\prime}$ slit achieves a resolution of $FWHM = 10~$\AA{}.
 Eight $900$~s exposures were obtained  with the mean time of 3.893~h after the burst. The spectra were averaged in the processing.
The average spectrum is reduced and calibrated against the spectra-photometric standard G191 B2B.
\bigskip

 The absorption lines detected in the GTC and BTA spectra (see Fig.~\ref{spec}) give  an
average redshift of $1.3202\pm0.0003$ and $1.319\pm0.003$, respectively. The equivalent widths of the spectral features are measured using the method
described by~\citet{Fynbo}. The results are presented
in Table~\ref{TabGTC}.

Using the methodology described by~\citet{2012A&A...548A..11D} we obtain a line strength parameter LSP $=-0.16\pm0.17$  for the spectrum shown in Fig.~\ref{spec} (GTC1)  implying that the absorption features in this
spectrum are stronger than 43 per cent of the GRB afterglows in their sample, very close
to the observed average of GRB afterglow lines of sight. We note the high
ionization of the environment shown by the high ratio between Mg~II
and Mg~I, indicating a possibly strong ionizing electromagnetic field in the host. The distribution of the rest-frame equivalent widths for the lines in the GRB afterglow is presented in Fig.~\ref{GTCew}.

Within the spectrum we also detect [O II] $\lambda$3727, $\lambda$3729 emission,
with a combined flux of $(3.3\pm0.2)\times10^{-17}$~erg~cm$^{-2}$~s$^{-1}$.  These forbidden lines enable estimates of the star formation rate in the host galaxy following~\citet{1998ARA&A..36..189K}, which we make in the following section.

\section{Discussion}
\subsection{The GRB environment}
Using the photometry data taken by GROND in 6 filters at $\sim$\,0.63 day (54370-54377s) after $T_0$ and contemporary X-ray observations by \textit{Swift}~XRT between 36390 and 127966 s, we build and fit the spectrum of the GRB. The \textit{Swift}~XRT spectrum is prepared at the \textit{Swift} XRT spectra repository~\citep{Evans2009}.
The XRT spectrum is rebinned using tool {\tt grppha} requiring at least 5 counts per bin, for the sake of using $\chi^2$ minimization during fitting. The optical data is converted by the task {\tt flx2xsp}. We use two xspec models: powerlaw and bknpower (broken power law) multiplied by absorption components zdust*tbabs*ztbabs. The resulting number  of d.o.f. is 13 and 12, for powerlaw and bknpower, respectively. For bknpower, we fix the difference between the photon indexes to 0.5,
which corresponds to the difference of spectral slopes in the Slow Cooling regime, lower and higher than the characteristic cooling frequency $\nu_c$ (see~\citealt{GranotSari2002,Racusin2009}). Optical fluxes are corrected for the extinction in the Galaxy using the NASA/IPAC Extragalactic Database Galactic Extinction calculator~\footnote{\url{http://ned.ipac.caltech.edu/forms/calculator.html}},
based on the work by~\citet{Schlafly2011}: $A_g=0.78$, $A_r=0.52$, $A_i=0.38$,  $A_z=0.28$, $A_J=0.16$, and $A_H=0.10$ mag.

\begin{figure*}
\center{\includegraphics[trim=0 0 0 0,clip, width=0.5\textwidth]{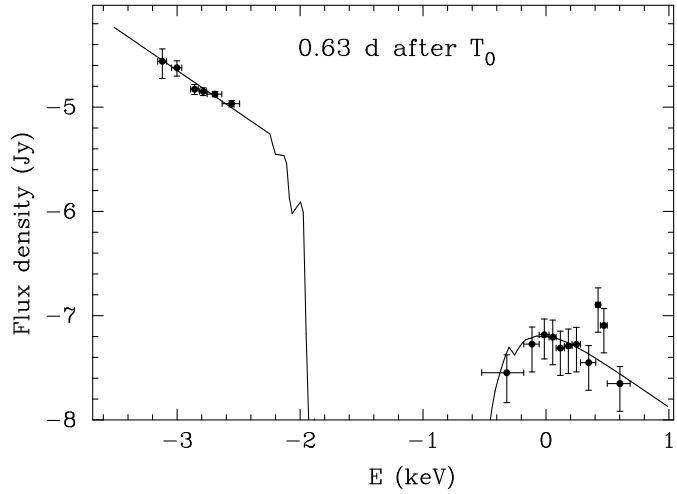}}
\caption{Spectrum of GRB~140801A compiled from the XRT data in
the interval  from about 36.4 to 128 ks and the GROND $g'r'i'z'JH$
data obtained at $\sim$\,0.63 d after the trigger, corrected
for the Galactic extinction. The solid line is the best-fitting absorbed power-law model,
whose parameters are given in the text.}
\label{fig.spec_GROND_XRT}
\end{figure*}

\begin{figure*}
\centering
\center{\includegraphics[width=0.8\textwidth]{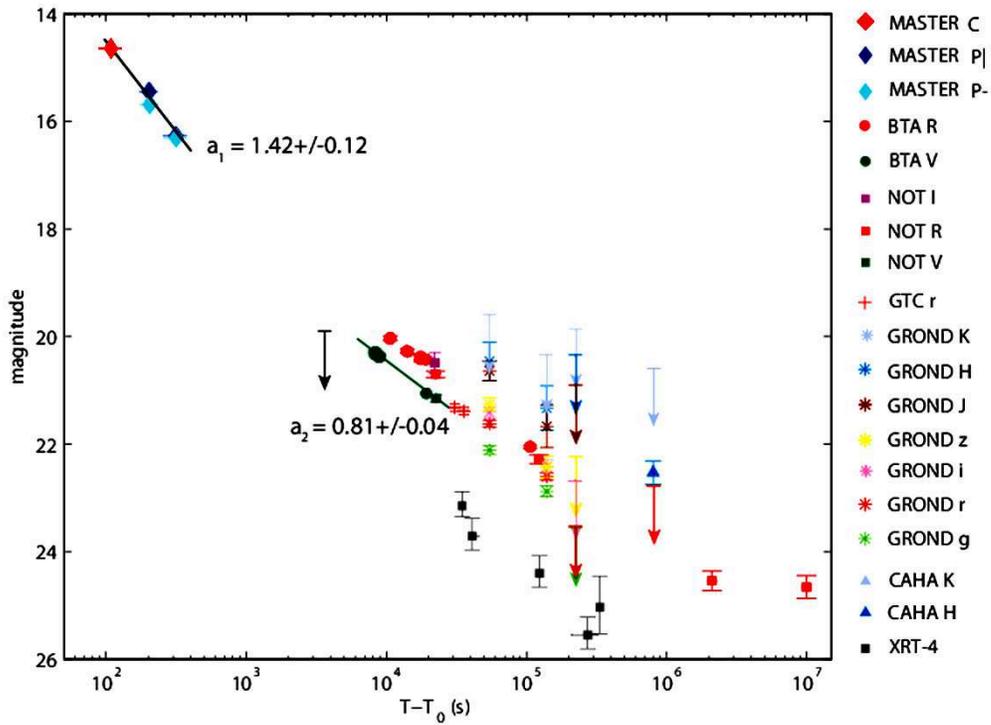}}
\caption{Combined light curve of GRB~140801A. The arrows designate upper limits. The OT decays as $t^{-1.42\pm0.04}$ during the MASTER observations, and then, after 2 hours, changes to  $t^{-0.81\pm0.04}$ in $V$ band assuming a simple power-law decline ($m = 2.5ax+b$, where $x = log(t)$). The XRT observations are shown in gray as $m(\mathrm{AB})-4$.}
\label{LC}
\end{figure*}

\begin{figure*}
 \centering
 \center{\includegraphics[width=0.8\linewidth]{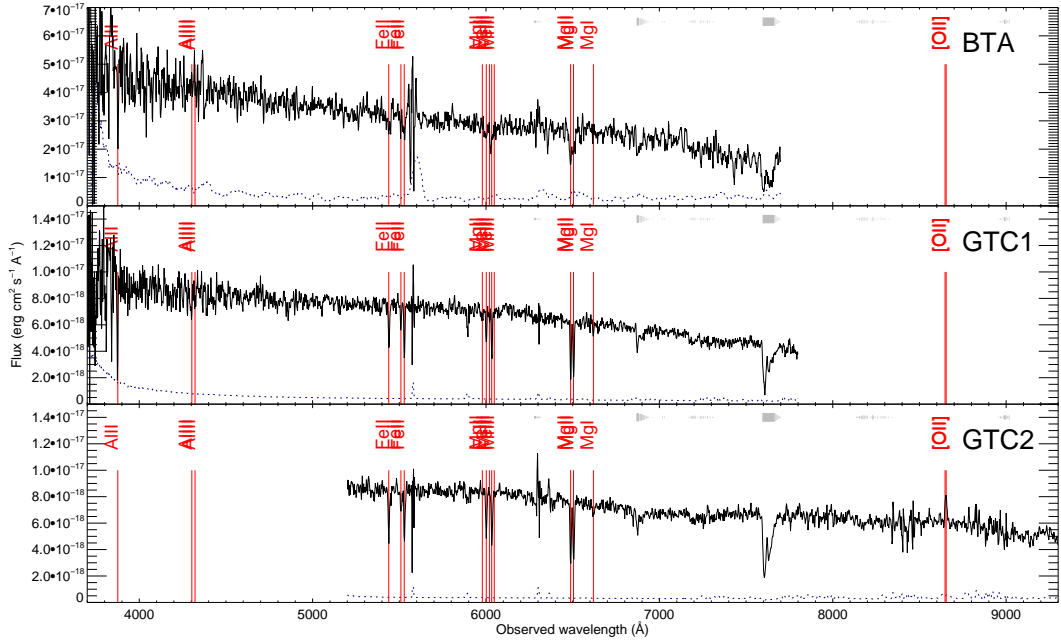}}
 \caption{The spectrum of the GRB\,140801A optical afterglow obtained with the 6-m BTA/Scorpio-I 3.893 h after the burst and with the 10.4-m GTC 8.987 and 10.356~h after the burst, respectively.
 The redshift determined from the lines is $z=1.32$~\citep{GCN16657,GCN16663}.}
 \label{spec}
 \end{figure*}

\begin{figure*}
\centering
\center{\includegraphics[width=1\linewidth]{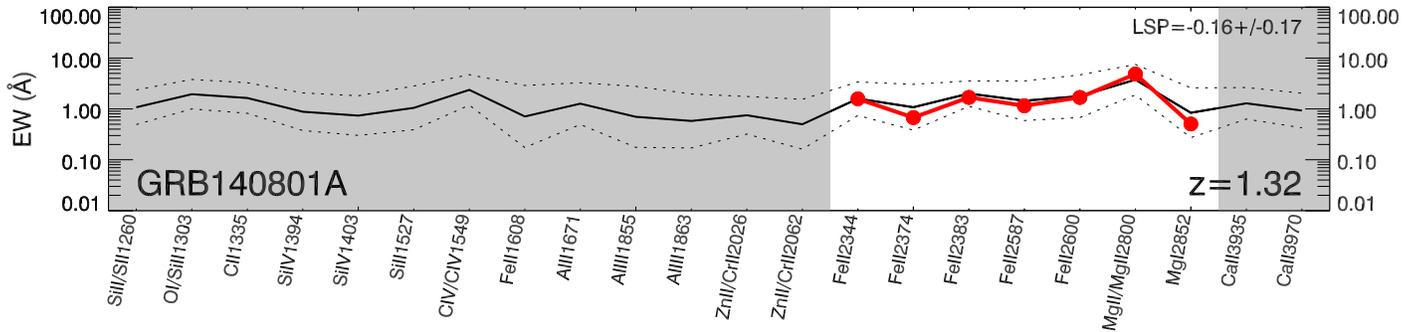}}
\caption{EW diagram: the bold line represents the  equivalent widths of the spectral features of GRB\,140801A. The continuous black line is the average of
the sample of GRBs (see Fig. 2 of~\citealt{2012A&A...548A..11D}), whereas the dotted lines indicate the standard deviations over and under this value. The shaded regions show regions of the spectra where there is no data.}
\label{GTCew}
\end{figure*}

We have tested three types of zdust model (extinction by the dust grains) with different values of the total-to selective-extinction $R_V$, according to~\citet{Pei1992}.
For the single power law and for every zdust model, the fit is satisfactory ($\chi^2/dof=10.2/13$)
with photon spectral index $\Gamma = 1.81^{+0.04}_{-0.03}$. All errors and limits reported
are for the 90~per cent confidence range. In Fig.~\ref{fig.spec_GROND_XRT}, the spectrum and fit are shown. For the broken power law with fixed
difference between the spectral slopes, we find satisfactory fits with $E_\mathrm{break}=0.01_{-0.008}^{+0.01}$~keV for the Milky Way (MW)-type zdust model ($\chi^2/dof=10.6/12$),
and  $E_\mathrm{break}=0.01_{-0.008}^{+0.02}$~keV for the Large Magellanic Cloud (LMC)-type zdust model ($\chi^2/dof=11.4/12$). For the Small Magellanic Cloud (SMC)-type, there is no acceptable $E_\mathrm{break}$ in the fitted spectral range. For MW- and LMC-type zdust, the lower photon index $\Gamma_\mathrm{lower} \simeq 1.47$.

For the single power-law spectrum,  the optical extinction in the host is negligible with upper limit $A_V^\mathrm{int}\lesssim 0.04$  for the MW and LMC dust absorption models and $A_V^\mathrm{int}\lesssim 0.03$ for the SMC model.
For  the broken power-law fits, we obtain higher upper limits on $A_V^\mathrm{int}\simeq 0.16$.
Intrinsic absorption of the soft X-rays at $z=1.32$ yields an upper limit
on the number of hydrogen atoms $N_\mathrm{H}^\mathrm{int}$, $2.4 \times 10^{21}$cm$^{-2}$ and $6.5 \times 10^{21}$cm$^{-2}$, for single power-law
and broken power-law, respectively. These are consistent with the upper limit reported for the late \textit{Swift}~XRT spectrum at the UK Swift Science Data Centre ($2\times 10^{21}$~cm$^{-2}$). The Galactic absorption column density
by H~I and H$_2$ is fixed at $N_\mathrm{Htot}^\mathrm{int}$ $1.74\times 10^{21}$~cm$^{-2}$~\citep{Willingale2013}.

Assuming the afterglow of GRB~140801A is in the slow cooling regime in the time-span $10^4$--$10^5$~s~\citep{Sari1998}, and that the
cooling frequency $\nu_c$ is at $\sim$\,0.01~keV, we can reconcile the
optical light curve slope and the spectral slopes found by the
spectral fitting below and above $\nu_c$ with the forward shock
spreading into a constant-density interstellar medium with a power-law index of the Lorentz factor distribution of the electrons of 1.9 (Table 1 of \citealt{Racusin2009}). For the optical
spectral slope $\beta = \Gamma_\mathrm{lower}-1 \simeq 0.47 $
the expected  light curve slope is $ \simeq 0.74$~\citep{Racusin2009}, which is marginally consistent with the slope found in Subsection 3.4 (see Fig.~\ref{LC}).

If the cooling frequency lies redward of the optical range,
the fit with $\beta_\mathrm{opt-Xray} \simeq 0.8$  can be also explained by the forward shock spreading into a constant-density
interstellar medium, requiring the index of the
power-law distribution of the electrons' Lorentz factors $1.6$~\citep{Racusin2009}, which would result in
a light curve slope
 $\simeq 0.9$.

We note that such shallow light curve
slopes are sometimes modeled by continuous energy injection~\citep{Zhang2006}.

Following a diagnostic proposed in~\citet{1998ARA&A..36..189K}, we correct the observed luminosity of the O~II line for extinction at  the wavelength where we see O~II ($3728\times(1+1.32) = 8649$~\AA) in the Galaxy $A_{8649}^{\mathrm{MW}} = 0.39$ and, optionally,  in the host $A_{3728}^{\mathrm{int}} \leq 1.54\times A^\mathrm{int}_V = 0.25$~\citep{Pei1992}.
For $D_l = 9.6$~Gpc, using the forbidden-line calibration for the star-formation rate by ~\citet{1998ARA&A..36..189K}, eq.~3 we obtain the values $7.3$ and $9.1$~M$_{\odot}$ yr$^{-1}$, for the boundary values $A^\mathrm{int}_V=0$ and $A^\mathrm{int}_V=0.16$, respectively.

\subsection{Amati and Ghirlanda relations}
From the high-energy analysis we found the isotropic rest-frame energy $E_\mathrm{iso} = 5.54_{-0.24}^{+0.26} \times 10^{52}$~erg  and the observed $E_\mathrm{peak}=$110--130~keV.
The peak of the energy distribution in the rest frame is at $E_\mathrm{p, rest} = (1+z)\times E_\mathrm{peak} = $255--302~keV. The $E_\mathrm{iso}$ and $E_\mathrm{p, rest}$ are close to the Amati correlation within the measurement accuracy~\citep{Amati2002,Nava}.

Due to an apparent absence of a light-curve break at times later than 10$^4$ s, we can infer only a lower limit on $t_\mathrm{break}$, which is $1.4\times 10^5$ s. Assuming the values of the efficiency of converting the ejecta energy into $\gamma$-rays $\eta_{\gamma} = 0.2$ and the number density of the ambient medium $n = 3 cm^{-3}$ --- we obtain a lower limit on the  half-opening angle of the jet $\theta=6.1$ deg and the collimation corrected energy $E_{\gamma} = 3.14\times 10^{50}$ erg. This value, together with $E_\mathrm{p, rest}$, agrees with the Ghirlanda relation: $E_\mathrm{p, rest}\simeq 480\times(E_{\gamma}/10^{51} \mathrm{erg})^{0.7}$~\citep{Ghirlanda2004}.

\subsection{Results of MASTER polarization measurements}
Results of  polarization measurements of GRB~100906A, GRB~110422A, and GRB~121011A  by MASTER have been reported by~\citet{Gorbovskoy2012,Pruzhinskaya2014}.
For GRB~100906A, from 0.01 to 0.19 h after $T_0$, it was found that the time-averaged dimensionless Stokes parameter was less than the standard deviation of that of the field stars (2 per cent for that case). GRB~110422A (from 0.02 to 0.51 h after $T_0$) and GRB~121011A (from 0.02 to 2.70 h after $T_0$)  have not shown dimensionless Stokes parameter greater than 2 per cent either. Individual measurements of the Stokes parameter were also around zero. Two MASTER telescopes are needed to observe simultaneously two Stokes parameters. As a single Stokes parameter is the lower estimate for the degree of the linear polarization $P_L$, we can estimate the 1-$\sigma$ upper limit on $P_L$. For 2 per cent uncertainty of the dimensionless Stokes parameter, these GRBs had $P_L\lesssim15$~per cent at  1-$\sigma$ level.
Thus, unfortunately,  polarization results for GRB~140801A, GRB~100906A, GRB~110422A, and GRB~121011A
cannot be used to support specific models of the reverse or forward shock with  polarization predictions.

\section{Conclusions}
In this paper we presented high-energy, photometric, and spectroscopic observations of GRB~140801A. Detailed studies of individual gamma-ray bursts in different wavelength and energy regimes, especially during the early stage of emission, are still essential.  GRB 140801A at $z=1.32$ could be classified as an intermediate gamma-ray burst by its duration, but it has a hardness
ratio characteristic for long GRBs~\citep{Horvath}. GRB~140801A is a typical GRB in many ways. It obeys the Amati correlation~\citep{Amati2002,Nava}.
A negligible spectral lag is not unknown for long GRBs, but
it is rather rare --- and more typical for short GRBs. Though this
burst seems to have a high peak luminosity, which would agree with
the small lag. Based on results by \citet{Wang2013}, we find that the optical afterglow of GRB~140801A lies
in the brightest 20\% of observed optical afterglows at 100~s after
the burst trigger.
The optical light curve behaviour can be described by an early, steep decline $t^{-1.42\pm0.04}$ which becomes slower ($t^{-0.81\pm0.04}$ in $V$ and $t^{-0.82\pm0.01}$ in $R$ bands) after approximately 2 hours. We conclude that there is no evidence of jet break in our optical data.

Spectral modeling of the optical and X-ray data around 0.63~d,
together with the derived slope of the optical emission, favour a model in which the emission is
generated by the forward shock spreading into a homogeneous
interstellar medium.

We have  demonstrated the possibility to rapidly detect the optical counterparts of events with a poorly defined localization. This capability is extremely important due to ongoing experiments with localization areas of more than 100 deg$^2$~\citep{Singer2014,Singer2015}. The discovery of optical counterparts of gravitational or radio alerts may prove to be crucial in shedding light on the physical nature of these events. However, this is possible only via a strong cooperation between space, ground-based and robotic observatories.

\section*{Acknowledgments}
MASTER Global Robotic Net is supported in part by the  Development Programm of Lomonosov Moscow State University.
This work was also supported in part by RFBR 15-02-07875 grant.

This work is also partially supported by the Russian Federation Ministry of Education and Science (agreement 14.B25.31.0010 and  government assignment 2014/51, project 1366) and by state order No. 3.615.2014/K in relation to scientific activity (design part).

The \textit{Fermi} GBM collaboration acknowledges support for GBM development, operations, and data analysis from NASA in the U.S.A and BMWi/DLR in Germany.  A.~Goldstein is supported by an appointment to the NASA Postdoctoral Program at MSFC, administered by Oak Ridge Associated Universities through a contract with NASA.

This work made use of data supplied by the UK Swift Science Data Centre at the University of Leicester. This research has made use of NASA's Astrophysics Data System.

The \textit{Fermi} GBM collaboration acknowledges support for GBM development, operations, and data analysis from NASA in the U.S.A and BMWi/DLR in Germany.  A.~Goldstein is supported by an appointment to the NASA Postdoctoral Program at MSFC, administered by Oak Ridge Associated Universities through a contract with NASA.

This work was supported in part by Russian Foundation of Fundamental Research, grant RFBR 14-02-31546, 14-02-91172.

This work made use of data supplied by the UK Swift Science Data Centre at the University of Leicester. This research has made use of NASA's Astrophysics Data System.

The Konus-WIND experiment is partially supported by a Russian Space Agency contract, RFBR grants 15-02-00532a and 13-02-12017 ofi-m.

Part of the funding for GROND (both hardware as well as personnel) was  generously granted from the Leibniz-Prize to Prof. G. Hasinger (DFG grant  HA 1850/28-1).

A.~J. Castro-Tirado acknowledges support from the Spanish Ministry Grant AYA 2012-39727-C03-01.

The data presented here were obtained in part with ALFOSC, which is provided by the Instituto de Astrofisica de Andalucia (IAA) under a joint agreement with the University of Copenhagen and NOTSA.

C.~Delvaux acknowledges support through EXTraS, funded from the European Union's Seventh Framework Programme for research, technological development and demonstration under grant agreement no 607452.

D.A.~Kann, S.~Klose, and S. Schmidl acknowledge support by DFG grants Kl 766/16-1 and 766/16-3.  In addition, S.~Schmidl acknowledges support by the Th\"uringer Ministerium  f\"ur Bildung, Wissenschaft und Kultur under FKZ 12010-514.

We acknowledge A.~Burenkov, V.~Vlasyuk and T.~Fatkhullin for the help in observations.
A.~Moskvitin, V.~Komarova and T.~Sokolova are partically supported by the Research Program OFN-17 of the Division of Physics, Russian Academy of Sciences. A.~Moskvitin is also partically supported by the grant MK-1699.2014.2 of the President of Russian Federation and by RFBR 14-32-50547.

S. Schulze acknowledges support from CONICYT-Chile FONDECYT 3140534, Basal-CATA PFB-06/2007, and Project IC120009 ``Millennium Institute of Astrophysics (MAS)´´ of Iniciativa Cient\'{\i}fica Milenio del Ministerio de Econom\'{\i}a, Fomento y Turismo.

\begin{table*}
 \centering
  \caption{Spectral parameters of GRB~140801A from \textit{Fermi} GBM and KW data.}
  \begin{tabular}{@{}lccccccccc@{}}
  \hline
  \hline
   Data     & Model &GBM-NaI&GBM-BGO& $\alpha$ & $\beta$ & $E_\mathrm{peak}$ & Flux\textsuperscript{1}& $\chi^2/$dof \\
     & & norm. factor&norm. factor&& & (keV) & $(\times 10^{-6}$ erg cm$^{-2}$s$^{-1})$            \\

 \hline

 \hline
\multicolumn{9}{ |c| }{Time-integrated spectrum (0--7.680 s)\textsuperscript{2}} \\
  \hline

KW & CPL &... & ... & $-0.44_{-0.17}^{+0.18}$ &... & $108_{-4}^{+5}$ & $1.50_{-0.07}^{+0.07}$ & 75/71\\
KW & Band & ... &... & $-0.44_{-0.15}^{+0.17}$ & $< -4.2$ & $108_{-4}^{+5}$ & $1.48_{-0.05}^{+0.07}$ & 75/70\\
GBM & CPL & 1 & $1.52_{-0.32}^{+0.34}$ & $-0.27_{-0.07}^{+0.07}$ &...  & $123_{-4}^{+4}$ & $1.44_{-0.03}^{+0.04}$ & 318/465\\
GBM & Band & 1 & $1.47_{-0.33}^{+0.35}$ & $-0.26_{-0.07}^{+0.08}$ & $< -3.5$ & $122_{-4}^{+4}$ & $1.47_{-0.06}^{+0.09}$ & 317/464\\
KW+GBM\textsuperscript{3}& CPL & $0.96_{-0.03}^{+0.03}$ & $1.64_{-0.34}^{+0.36}$ & $-0.23_{-0.06}^{+0.06}$ &... & $117_{-3}^{+3}$ & $1.46_{-0.05}^{+0.05}$ & 421/538\\
KW+GBM\textsuperscript{3}& Band & $0.96_{-0.03}^{+0.03}$ & $1.64_{-0.34}^{+0.36}$ & $-0.23_{-0.06}^{+0.07}$ & $< -4.0$ & $117_{-3}^{+3}$ & $1.46_{-0.05}^{+0.07}$ & 420/537\\

\hline
\multicolumn{9}{ |c| }{Spectrum covering the peak count rate (0--0.256 s)\textsuperscript{2}} \\
\hline
KW & CPL & ... & ... & $+0.39_{-0.56}^{+0.69}$ & ... & $121_{-12}^{+13}$ & $3.20_{-0.36}^{+0.38}$ & 27/24\\
KW & Band & ... & ...& $+0.49_{-0.65}^{+0.94}$ & $< -3.2$ & $119_{-15}^{+15}$ & $3.25_{-0.40}^{+0.46}$ & 27/23\\
GBM & CPL & 1 & $1.26_{-0.64}^{+0.77}$ & $-0.02_{-0.23}^{+0.25}$ & ... & $140_{-12}^{+15}$ & $3.36_{-0.25}^{+0.28}$ & 86/128\\
GBM & Band & 1 & $1.19_{-0.65}^{+0.81}$ & $+0.02_{-0.25}^{+0.34}$ & $< -2.5$ & $136_{-19}^{+17}$ & $3.55_{-0.43}^{+1.32}$ & 86/127\\
KW+GBM\textsuperscript{3}& CPL & $0.96_{-0.09}^{+0.11}$ & $1.42_{-0.71}^{+0.82}$ & $+0.11_{-0.20}^{+0.22}$ & ... & $131_{-8}^{+10}$ & $3.36_{-0.33}^{+0.33}$ & 116/154\\
KW+GBM\textsuperscript{3}& Band & $0.96_{-0.09}^{+0.11}$ & $1.42_{-0.71}^{+0.82}$ & $+0.11_{-0.20}^{+0.23}$ & $< -3.4$ & $131_{-10}^{+9}$ & $3.35_{-0.17}^{+0.45}$ & 116/153\\
\hline
\multicolumn{7}{l}{\textsuperscript{1}\footnotesize{In the 1~keV -- 10~MeV range.} }\\
\multicolumn{7}{l}{\textsuperscript{2}\footnotesize{Since the KW trigger time.}}\\
\multicolumn{7}{l}{\textsuperscript{3}\footnotesize{The KW normalization factor is fixed to the unity.}}\\
\multicolumn{7}{l}{\footnotesize{Note. --- All the quoted errors are at the 90 per cent confidence level.}}
\end{tabular}
\label{fittab}\\
\end{table*}

\begin{table*}
\centering
\begin{minipage}{125mm}
  \caption{Photometric observations of the  GRB~140801A afterglow. MASTER $C$ band is the white light that approximately corresponds to 0.2$B+0.8R$. $P|$, $P$-- are polarizers oriented at 0$^\circ$ and 90$^\circ$ to the celestial equator, respectively.  $T_{start}$~is the time of the beginning of exposure, $T-T_0$~is the time between the middle of exposure and \textit{Fermi} GBM trigger time $T_0=18$:59:53.26~UT. The magnitudes are in the AB system and not corrected for the Galactic extinction.}
\begin{scriptsize}
\begin{tabular}{@{}cclcccc@{}}
\hline
Band          &$T_{start}$ (UT)                         & $T-T_0$   &Exposure (s)      &Magnitude	&Err. mag\\
 \hline
\multicolumn{6}{ |c| }{0.4-m MASTER Tunka} \\
 \hline
$C$                &2014-08-01T19:01:31.89		&108.6 (s)&20	                     &14.64	&0.07\\
$P|$	         &2014-08-01T19:03:01.01		&202.8         &30	                     &15.45	&0.15\\
$P$--	         &2014-08-01T19:03:02.86		&204.6         &30	                     &15.69	&0.11\\
$P|$	         &2014-08-01T19:04:38.38		&315.1         &60	                     &16.27	&0.07\\
$P$--	         &2014-08-01T19:04:38.96		&315.7         &60	                     &16.31	&0.08\\
$C$ 	         &2014-08-01T19:57:26.72		&3631.1       &3$\times$60	&$>$19.82\textsuperscript{1}  &	              \\
\hline
\multicolumn{6}{ |c| }{6-m BTA} \\
 \hline
$V$                &2014-08-01T21:16:04.50            &2.2726 (h)	&20                          &20.30         &0.05\\
$V$                &2014-08-01T21:18:16.82            &2.3093	&20                          &20.34         &0.05\\
$V$                &2014-08-01T21:21:10.34            &2.3575	&20                          &20.30         &0.05\\
$V$                &2014-08-01T21:27:16.21            &2.4605	&30                          &20.39         &0.04\\
$V$                &2014-08-01T21:30:44.84            &2.5185	&30                          &20.37         &0.04\\
$R_c$               &2014-08-01T21:56:27.22            &2.9455 	&20			&20.04         &0.04\\
$R_c$               &2014-08-01T21:57:56.49            &2.9703 	&20			&20.03         &0.04\\
$R_c$               &2014-08-01T22:53:25.75            &3.8951     &20			&20.27         &0.04\\
$R_c$               &2014-08-01T22:56:11.20            &3.9411	&20			&20.28         &0.04\\
$R_c$               &2014-08-01T23:50:27.04            &4.8455	&20			&20.42         &0.05\\
$R_c$               &2014-08-01T23:52:03.02            &4.8722	&20			&20.38         &0.05\\
$R_c$               &2014-08-01T23:54:26.23            &4.9119	&20			&20.38         &0.05\\
$R_c$               &2014-08-01T23:56:38.44            &4.9487	&20			&20.40         &0.05\\
$R_c$               &2014-08-02T00:18:51.70            &5.3204	&30			&20.43         &0.05\\
$V$                &2014-08-02T00:21:03.02            &5.3610	&60                          &21.06         &0.03\\
$Rc$               &2014-08-03T00:05:30.90            &29.4552	&5$\times$180 	&22.05         &0.03\\
\hline
\multicolumn{6}{ |c| }{2.5-m NOT} \\
 \hline
$I$                 &2014-08-02T01:07:22.96            &6.1527 (h)	&200			&20.49        &0.19\\
$R$                 &2014-08-02T01:11:31.48            &6.2217	&200			&20.70         &0.06\\
$V$                 &2014-08-02T01:15:37.08             &6.2900	&200			&21.15         &0.06\\
$R$                 &2014-08-03T04:59:18.39            &34.0320	&300 		           &22.28        &0.08\\
$R$                 &2014-08-11T03:39:12.24            &224.9053	&6$\times$300 	&$>$22.77\textsuperscript{1}      &                                     \\
$R$                 &2014-08-26T03:27:54.59            &584.9670	&6$\times$600 	&24.54         &0.18\\
$R$                 &2014-11-25T01:53:20.50            &2767.5576	&8$\times$600 	&24.66         &0.21\\
\hline
\multicolumn{6}{ |c| }{10.4-m GTC} \\
 \hline
$r$                  &2014-08-02T03:32:16.68            &8.5426 (h)	&20			&21.32         &0.07\\
$r$                  &2014-08-02T04:54:46.23            &9.9175	&20			&21.39         &0.07\\
\hline
\multicolumn{6}{ |c| }{2.2-m GROND} \\
 \hline
$J$                  &2014-08-02T09:44:09.88             &15.1048 (h)&144$\times$10 	                     &20.64	&0.18\\		
$H$                  &2014-08-02T09:44:09.88            &15.1048	&144$\times$10	                     &20.46	&0.36\\		
$K$                  &2014-08-02T09:44:09.88            &15.1048	&144$\times$10	                     &$>$19.60\textsuperscript{1}	&          \\		
$g$                  &2014-08-02T09:44:10.69            &15.1027	&24$\times$66                    &22.11	&0.08\\
$r$                  &2014-08-02T09:44:10.69             &15.1027	&24$\times$66                     &21.62	&0.06\\
$i$                  &2014-08-02T09:44:10.69             &15.1027	&24$\times$66                     &21.41	&0.09\\
$z$                  &2014-08-02T09:44:10.69             &15.1027	&24$\times$66                     &21.26	&0.12\\	
$J$                  &2014-08-03T09:02:13.31             &38.7713	&300$\times$10	                     &21.67	&0.40\\
$H$                  &2014-08-03T09:02:13.31            &38.7713	&300$\times$10	                     &21.33	&0.41\\
$K$                  &2014-08-03T09:02:13.31            &38.7713	&300$\times$10	                     &$>$20.38\textsuperscript{1}	&          \\				
$g$                  &2014-08-03T09:02:13.36            &38.7240	&24$\times$115                     &22.88	&0.10\\
$r$                  &2014-08-03T09:02:13.36             &38.7240	&24$\times$115                     &22.60	&0.07\\
$i$                  &2014-08-03T09:02:13.36             &38.7240	&24$\times$115                    &22.42	&0.12\\
$z$                  &2014-08-03T09:02:13.36             &38.7240	&24$\times$115                     &22.42	&0.20\\
$J$                  &2014-08-04T09:07:30.45            &62.7760	&288$\times$10                     &$>$20.90\textsuperscript{1}  &         \\
$H$                  &2014-08-04T09:07:30.45           &62.7760	&288$\times$10                     &$>$20.35\textsuperscript{1}  &         \\
$K$                  &2014-08-04T09:07:30.45           &62.7760	&288$\times$10                     &$>$19.91\textsuperscript{1}  &          \\
$g$                  &2014-08-04T09:07:31.33           &62.7689	&24$\times$115                    &$>$23.54\textsuperscript{1}  &         \\
$r$                  &2014-08-04T09:07:31.33            &62.7689	&24$\times$115                    &$>$23.52\textsuperscript{1}  &         \\
$i$                  &2014-08-04T09:07:31.33            &62.7689	&24$\times$115                    &$>$22.66\textsuperscript{1}  &         \\
$z$                  &2014-08-04T09:07:31.33            &62.7689	&24$\times$115                    &$>$22.26\textsuperscript{1}  &         \\
\hline
\multicolumn{6}{ |c| }{3.5-m CAHA} \\
 \hline
$H$                  &2014-08-11T01:17:15.98          &222.2980 (h) &56$\times$60			 &22.53         &0.21\\
$K$                  &2014-08-11T03:45:13.79          &224.7640     &56$\times$60                    &$>$20.60\textsuperscript{1} &                                 \\
\hline
\multicolumn{1}{l}{\textsuperscript{1}\footnotesize{3-$\sigma$ upper limit.}}
\end{tabular}
\label{TabPhot}\\
\end{scriptsize}
\end{minipage}
\end{table*}

\begin{table*}
 \centering
  \caption{ The equivalent widths of the spectral features of GRB~140801A in the spectra obtained by GTC.}
  \begin{tabular}{@{}llllll@{}}
  \hline
  \hline
  \multicolumn{4}{|c|}{$t=8.987$ h (R1000B)} \\
  \hline
Wavelength ({\AA}) & Feature            & Redshift & EW  ({\AA})               \\
 \hline
 5440.548 &       FeII 2344.21 &  1.3202 &   3.665$\pm$ 0.352 \\
 5509.290 &       FeII 2374.46 &  1.3202 &   1.569$\pm$ 0.263 \\
 5529.494 &       FeII 2383.79 &  1.3202 &   3.909$\pm$ 0.281 \\
 6001.622 &       FeII 2586.65 &  1.3202 &   2.679$\pm$ 0.258 \\
 6033.697 &       FeII 2600.17 &  1.3202 &   3.901$\pm$ 0.299 \\
 6488.668 &       MgII 2796.35 &  1.3202 &   5.551$\pm$ 0.304 \\
 6505.847 &       MgII 2803.53 &  1.3202 &   5.690$\pm$ 0.366 \\
 6619.048 &       MgI  2852.96 &  1.3202 &   1.175$\pm$ 0.238 \\
   \hline
  \multicolumn{4}{|c|}{$t=10.356$ h (R1000R)} \\
  \hline
Wavelength ({\AA}) & Feature            & Redshift & EW  ({\AA})               \\
  \hline
  5440.292 &       FeII 2344.21 &  1.3202 &   3.729$\pm$ 0.401 \\
  5509.485 &       FeII 2374.46 &  1.3202 &   1.778$\pm$ 0.350 \\
  5529.231 &       FeII 2383.79 &  1.3202 &   3.898$\pm$ 0.369 \\
  6002.714 &       FeII 2586.65 &  1.3202 &   2.816$\pm$ 0.320 \\
  6033.836 &       FeII 2600.17 &  1.3202 &   3.936$\pm$ 0.394 \\
  6488.410 &       MgII 2796.35 &  1.3202 &   5.281$\pm$ 0.387 \\
  6504.886 &       MgII 2803.53 &  1.3202 &   5.699$\pm$ 0.438 \\
  6619.982 &        MgI 2852.96 &  1.3202 &   1.152$\pm$ 0.332 \\

\hline
\end{tabular}
\label{TabGTC}\\
\end{table*}

\end{document}